\documentstyle[aps,epsf,multicol]{revtex}

\title{Quasispecies evolution on a fitness landscape with a fluctuating peak}

\author{Martin Nilsson}

\address{Institute of Theoretical Physics,
Chalmers University of Technology and G\"{o}teborg University, 
S-412 96 G\"{o}teborg, Sweden {\tt martin@fy.chalmers.se}}

\author{Nigel Snoad}
\address{Santa Fe Institute, 1399 Hyde Park Road, Santa Fe, New Mexico 87501 USA \\
The Australian National University, ACT 0200, Australia {\tt nigel@santafe.edu}}
\date{\today}

\begin{document}

\maketitle

\begin{abstract}

A quasispecies evolving on a fitness landscape with
a single peak of fluctuating height is studied. 
In the approximation that back mutations
can be ignored, the rate equations can be solved analytically.
It is shown that the error threshold on this class of dynamic 
landscapes is defined by the time average of the selection pressure. 
In the case of a
periodically fluctuating fitness peak we also study the phase-shift
and response amplitude of the previously documented low-pass filter
effect. The special case of a small harmonic fluctuation is treated
analytically.

\end{abstract}

\begin{multicols}{2}
\narrowtext

\section{Introduction}

Ever since Eigen's work on replicating molecules in
1971~\cite{Eigen71}, the quasispecies concept has proven to be a
very fruitful way of modeling the fundamental behavior of evolution. A
quasispecies is an equilibrium distribution of closely related gene
sequences, localized around one or a few sequences of high fitness.
The kinetics of these simple systems has been studied in great detail,
and the formulation has allowed many of the techniques of statistical
physics to be applied to replicator and evolutionary systems, see for
instance~\cite{Eigen71,Eigen77,Schuster86,Schuster85,Leuthausser86,Tarazona92,Swetina88,NS89}.

The appearance in these models of an error-threshold (or
error-catastrophe) as an upper bound on the mutation rate above
which no effective selection can occur, has important implications
for biological systems. In particular it places limits on the
maintainable amounts of genetic
information~\cite{Eigen71,Eigen77} which puts restrictions on
theories for the origins of life.

Until now studies of quasispecies have mainly focused on static
fitness landscapes. However, many organisms in nature live in a quickly
changing environment. In this paper we will study
how a population responds to changes in the fitness landscape. More
precisely we will study the population dynamics on a fluctuating
single peaked fitness landscape. Since the full theory turns out to be
impossible to solve analytically, we introduce a simple
approximation that makes the rate equations analytically
tractable. The expression for the error threshold is then obtained
from the expression in the static case by replacing the 
height of the fitness peak by the time
average of the height of the fluctuating peak.
We also study how the phase-shift between fitness
oscillations and population dynamics depends on the frequency
in the case of a small harmonic fluctuation.

\section{Quasispecies in dynamic environments}

A quasispecies consists of a population of self-replicating genomes,
where each genome is
represented by a sequence of bases $s_k$, $\left( s_1 s_2 \cdots
s_{\nu} \right)$. We assume that the bases are binary, $s_k \in \{ 1, 0 \}$
and that all sequences have equal length $\nu$.  Every genome is then
given by a binary string $\left(011001 \cdots \right)$, which also can be
represented by an integer $k=\sum_j s_j2^j$ ($0 \leq k < 2^{\nu}$).

To describe how mutations affect a population we define $W_k ^l$ as
the probability that replication of genome $l$ gives genome $k$ as
offspring. We only
consider point mutations, which conserve the genome length $\nu$.

We assume that the point mutation rate $\mu = 1 - q$ (where $q$ is the
copying accuracy per base) is constant in time and independent of the
position in the genome.  We can then write down an explicit expression for
$W_k ^l$ in terms of the copying fidelity:

\begin{eqnarray}
    W_k ^l & = & \mu ^{h_{k l}} q ^{\nu - h_{k l}} = q ^{\nu} 
    \left( \frac{1-q}{q} \right) ^{h_{k l}} \label{eq1}
\end{eqnarray}
where $h_{k l}$ is the Hamming distance between genomes
$k$ and $l$. The Hamming 
distance $h_{k l}$ is defined as the number of 
positions where the genomes $k$ and $l$ differ.

The equations that describe the dynamics of the population take a
relatively simple form. Let $x_k$ denote the relative concentration
and $A_k (t)$ the time-dependent fitness of genome $k$. We then obtain
the following rate equations:

\begin{eqnarray}
     \dot{x}_k (t)  & = & \sum _l W_k ^l A_l (t) x_l (t) - e(t) x_k (t) 
             \label{eq2}
\end{eqnarray}
where $e (t) = \sum _l A _l (t) x_l (t)$, and the dot denotes a time
derivative.  The second term ensures the total normalization of the
population ($\sum _l x_l (t) = 1$) so that $x_k (t)$ describe
relative concentrations.

In the classical theory introduced by Eigen
and coworkers \cite{Eigen71,ES79,EMcCS89}, the fitness landscape is
static. The rate equations (\ref{eq2}) can then be solved analytically
by introducing a change of coordinates that makes them linear,
and then solving the eigenvalue system for the matrix
$ W_k ^l A_l$. 
The equilibrium distribution is given by the
eigenvector corresponding to the largest eigenvalue. 

If the fitness
landscape is time-dependent, this method cannot be applied. A
time-ordering problem occurs when we define exponentials of time-dependent
matrices, since in general the matrix
$ W_k ^l A_l (t)$ does not commute with itself at different
points in time. Later in this paper we make a simple approximation
that makes the rate equations one-dimensional; time-ordering
is then no longer necessary.

Much of the work on quasispecies has focused on fitness landscapes
with one gene sequence (the master sequence) with superior fitness,
$\sigma$, compared to all other sequences. These are viewed as a
background with fitness $1$. These landscapes are referred to as single peaked
landscapes. The master sequence is denoted $x_0$. In this paper we
focus on single peaked landscapes where the height of the fitness peak 
is time-dependent. The fitness landscape is then given by

\begin{eqnarray}
    A_k (t) & = & \left\{  \begin{array}{ll} \sigma (t) & \mbox{if } k=0   \\
                                        1 & \mbox{otherwise} \end{array} \right.
             \label{eq3}
\end{eqnarray}
This class of time-dependent landscapes was studied by Wilke and
co-workers \cite{Wilke99b,Wilke..99d}. They investigated the
behavior of a periodically fluctuating single peak landscape 
by numerically integrating the dynamics to find
the limit cycle of the concentrations for a full period. 

\begin{figure}
\centering
\leavevmode
\epsfxsize = .9 \columnwidth
\epsfbox{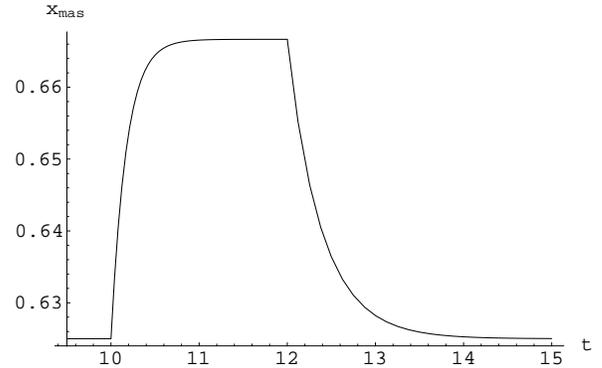}
\caption{The concentration of the master sequence 
when the fitness peak makes a sudden jump. The fitness is
given by $\sigma = 10$ when $t \in \left[ 10 , 12 \right]$, $\sigma =
5$ otherwise. The genomic copying fidelity is given by $Q = 0.7$. }
\label{hat}
\end{figure}

Fig.~\ref{hat} shows how the concentration of the master sequence
responds to a sudden, sharp jump in its fitness. When the fitness
changes it takes some time for the population to reach the new
equilibrium. It is this delay that causes a phase shift between a
periodically changing fitness function and the response in the
concentrations. The relaxation time of the population to the
appropriate equilibrium distribution depends on both the fitness
values of the landscape and the mutation rate. For extremely slow and
smooth changes in the fitness the population will effectively reach
equilibrium at every point in time. Thus the continued existence of a
quasispecies will depend on the local dynamics of the landscape. When
the landscape changes quickly the population will fail to follow the
changes adequately and thus responds to the landscape dynamics in
a way that is typical of a low pass filter. The following section
examines the fluctuating single peak landscape in some detail. 
In particular, we introduce an approximation that lets us find
an analytic form for the relaxation time of the
population, and the phase lag it introduces in a periodic landscape.

\section{Approximate quasispecies dynamics}

We now introduce a simple approximation of
the model presented above.
In this approximation we can solve
the rate equations and find an expression for
the concentration of the master sequence $x_0 (t)$. 
In the limit of long chain-length
($\nu \gg 1$)  we can
neglect back-mutations from the background to the master sequence.
This gives a simplified one-dimensional version of the rate equation
of the following form:

\begin{eqnarray}
    \dot{x}_0 (t) & = & Q \sigma (t) x_0 (t) - e (t) x_0 (t)
\label{approx}
\end{eqnarray}
where $Q = q^{\nu}$ is the copying fidelity of the whole genome and $e(t) = 
( \sigma (t) -1) x_0 (t) + 1$. 

\begin{figure}
\centering
\leavevmode
\epsfxsize = .9 \columnwidth
\epsfbox{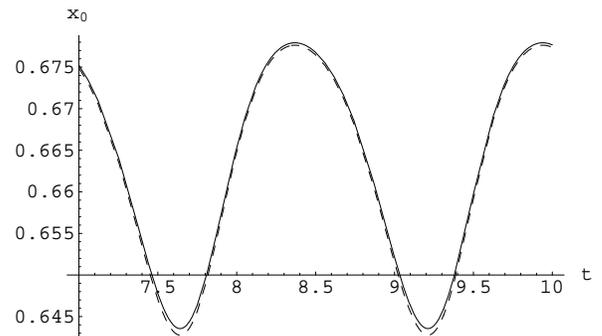}
\caption{The concentration of the master sequence calculated using the approximation~\ref{approx}
(dashed) and by numerically solving the rate equations~\ref{eq2} (solid). The fitness is
given by $\sigma = 10 + 5 \sin (4 t)$. The genomic copying fidelity is given by $Q = 0.7$
and the genome length $\nu = 25$. }
\label{num}
\end{figure}

Fig.~\ref{num} compares the concentration of the master sequence calculated by
solving approximation~\ref{approx} and by numerically integrating the 
full rate equation~\ref{eq2}. The figure shows that the approximation is quite accurate.

Since this equation is one-dimensional there is no time-ordering
problem and it can be solved analytically for non-periodic
peak fluctuations. Equation~\ref{approx} can be transformed to a linear form
by introducing a new variable $y (t) = \frac{Q - x_0 (t)}{(1-Q) x_0 (t)}$. This
gives

\begin{eqnarray}
	\dot{y} (t) & = & 1 - ( Q \sigma (t) -1) y (t) 
\label{eq4}
\end{eqnarray}
which can be solved. Substituting back gives the concentration of the master sequence

\begin{eqnarray}
	x_0 (t) & = & x_0 \frac{e^{\int _0 ^t (Q \sigma (s) - 1) d s}}
	{1+x_0 \int _0 ^t e^{\int _0 ^s (Q \sigma (u) -1) d u } (\sigma (s) -1) d s}
\label{eq5}
\end{eqnarray}

Since we are only interested in the long time behavior of the system we can ignore
transients carrying memory from initial values. Assuming $e^{\int _0 ^t (Q \sigma (s) - 1) d s} \gg 1$
gives 

\begin{eqnarray}
	x_0 (t) & = & \frac{Q}
	{1+(1-Q) \int _0 ^t e^{- \int _s ^t (Q \sigma (u) -1) d u } d s}
\label{eq6}
\end{eqnarray}
This is a generalization of the static expression for the asymptotic
concentration:
\begin{eqnarray}	
	x_0^s & = & \frac{Q \sigma -1}{\sigma -1}
\label{eq6.5}
\end{eqnarray}

On a static single peaked fitness landscape there is a phase
transition in the concentration distribution when the copying
fidelity decreases below a critical
value~\cite{Eigen77,M-SS95}. At high mutation rate the
selective advantage of the master sequence due to its superior fitness
is no longer strong enough for the gene sequences to be localized in
sequence space. Instead they diffuse over the entire sequence space,
and the distribution becomes approximately uniform. This is generally
referred to as the error catastrophe or error threshold and is one of
the main implications of the original quasispecies model. By making
the same approximation as above, i.e. assuming no back-mutations onto the
master sequence, the static landscape error threshold can be shown to occur when $Q =
\frac{1}{\sigma}$. In other words, the transition occurs when the 
selective advantage of the
master-sequence no longer is  able to compensate for the loss
of offspring due to mutations.  This can also be seen from
Eq.~\ref{eq6.5} which defines the stationary distribution of the
master sequence in the static case.

One has to be careful when discussing the error threshold on a fluctuating peak. 
The fitness can, for example,
slowly move from being strong enough to localize the population around
the peak, to beibg so weak that the population delocalizes, and then back
again. If we however consider an average over a time scale much longer
than the fluctuation time of the fitness peak, a sensible definition
of the error threshold can be made based on the average concentration
of the master sequence. The time average of the concentrations can be
found by rewriting equation~\ref{approx} as differentials
\begin{eqnarray}
	\int \frac{dx_0}{x_0} & = 
	& \int ( Q \sigma (t) - 1 - (\sigma (t) -1) x_0 (t)) d t 
\label{eq7}
\end{eqnarray}

The concentration of the master-sequence is positive. The left hand side of Eq.~\ref{eq7}
is therefore positive and the last term in the integral, $-(\sigma (t) -1) x_0 (t)$,
on the left hand side is negative. This implies that for $x_0 (t)$ to be positive as 
time goes to infinity, we must assume $\int \left( Q \sigma (t) -1 \right) d
t >0$. The fluctuating time dependent equivalent to the static error threshold is
therefore given by

\begin{eqnarray}\label{eq8}
	Q_{fc} & = & \frac{1}{\langle \sigma \rangle_t}
\label{eq7.5}
\end{eqnarray}
This shows that the error threshold on a fluctuating
fitness peak is determined by the time average of the fitness, if the 
fluctuations are fast compared to the response time of the population.

Eq.~\ref{eq6} indicates that the response time of the system is
approximately given by $( Q \sigma (t) -1) ^{-1}$, i.e. the relative growth
of the mastersequence compared to the background. For the time
average mentioned above to be an interesting parameter the
fluctuations of the fitness peak must therefore be faster than this
response time; only for this kind of environmental dynamics is it
sensible to talk in terms of the average concentration of the
master-sequence. Thus if the fluctuations occur on a time-scale faster
than the response-time of the quasispecies, then the error-threshold
is defined by Eq.~\ref{eq7.5}. For extremely slow changes the system will
effectively be in equilibrium around the current value of the
fitness. For slightly faster changes the response of the population
will lag somewhat behind the changes in selective environment.  In
these cases it is more interesting to study the minimal concentration
of the master sequence, which occurs when the fitness peak has a
minimum (as we shall see later the phase-shift decreases when the
fluctuation frequency decreases).

\begin{figure}
\centering
\leavevmode
\epsfxsize = .9 \columnwidth
\epsfbox{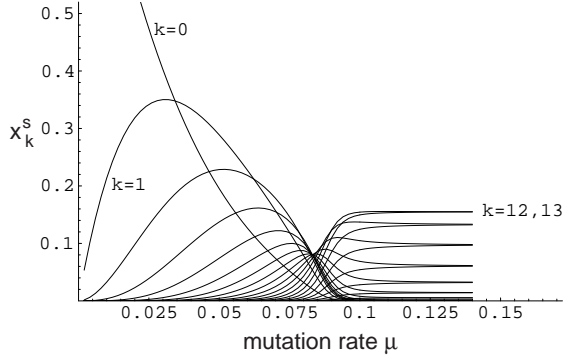}
\caption{The time-averaged quasispecies distribution is shown as a 
function of the error rate $\mu=q-1$. The figure shows the numerical solution to the full 
rate equations.
The fitness peak is  defined as $\sigma (t) = 10 + 5 \sin (t)$ and the genome length is $\nu=25$. The 
error threshold is located at $\mu \approx 0.085$, corresponding to $Q_{fc} = 0.109$ which can be compared to the approximate value $Q_{fc} = 0.1$ predicted
by Eq.~\ref{eq8}.}
\label{error}
\end{figure}

When the full replicator equations for a rapidly fluctuating peak are
numerically integrated, the time-averaged
quasispecies distribution displays an error catastrophe at high
error rates $\mu = 1-q$. In figure~\ref{error} the fitness
peak fluctuates periodically with $\sigma (t) = 10 + 5 \sin (t)$. The
average fitness is given by $\langle \sigma \rangle_t = 10$ and the genome length
$\nu=25$ and thus Eq.~\ref{eq8} predicts the error-threshold to occur
at $\mu = 0.088$, which agrees with the value found by numerically
integrating the equations of motion directly.  The analysis
in this section demonstrates that by making the error tail
approximation and reducing the dynamics to one-dimensional form, an
analytic form exists for the error-threshold on fast moving
landscapes. This one-dimensional formulation removes the need to
time-order the changes in selective advantage of the landscape.  This
allows the integrals for the time history of the master-sequence
concentration to be solved explicitly in equation~\ref{eq5}.

\section{Phase-shifts on periodic landscapes}

To study how the master sequence responds to changes in the height of
the fitness peak it is convenient to assume that the fluctuations are
periodic. It then makes sense to speak of the amplitude of the
oscillations in concentration and of the phase-shift between the
concentration and the fitness. It is intuitively clear that when the
fitness peak is oscillating slowly (compared to the response time $(Q
\sigma (t) -1)^{-1}$) there will be a very small phase-shift; the
population will have time to reach an equilibrium about every value of
$\sigma(t)$. The amplitude of changes in the master-sequence concentration
will, for the same reason, be as large as possible. This result,
together with the time-averaging effect found in the previous section,
indicates that the population responds to the driving of the
environment with a low pass filter effect. In one-dimensional
population genetic models this phenomenon has been noted for some time
\cite{Ishii..89,Charlesworth93,LS96}. Wilke
\textit{et al.}~\cite{Wilke99b} demonstrated via simulations that the same
filtering occurred to quasispecies evolution on a periodically
fluctuating single peak. Noting that the maxima and minima in
concentration occurs when $\dot{x}_0 = 0$, we can find a relation
between the phase-shift (between the concentration and fitness
fluctuations), and the amplitude of the fitness fluctuations. Let
$t_{xmax}$ be the time when the concentration has a maximum. Similarly
the fitness is at a maximum at time $t_{\sigma max}$. Thus the
phase-shift between the two is $\delta = t_{xmax} - t_{\sigma
max}$. From equation~\ref{approx} the condition for the maximum value
of $x_0$ during a full cycle can be derived
\begin{eqnarray}
	\max _t ( x_{0} ( t )) & = & \frac{Q \sigma (t_{\sigma max} + \delta ) - 1}
	{\sigma (t_{\sigma max } + t_{\delta}) - 1}
\label{eq9}
\end{eqnarray}

\begin{figure}[!t]
\centering
\leavevmode
\epsfxsize = 0.9 \columnwidth
\epsfbox{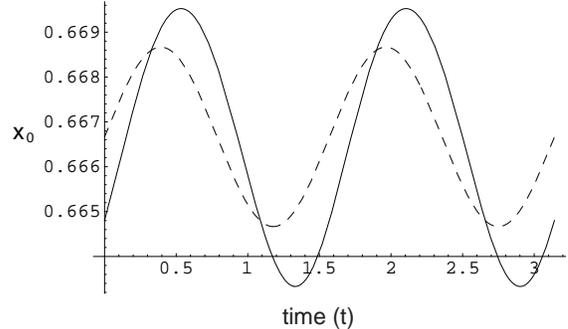}
\caption{\small  Thhe response in concentration of the master 
sequence (solid line) as the fitness peak oscillates according to $ \sigma (t) = 10
+ \sin ( 4 t)$. The genomic copying fidelity is $Q = 0.7$. The dashed
line shows $\sigma (t)$, scaled to fit in the plot. Note the
phase-shift between the fitness function and the concentration
response.}
\label{fig:harm}
\end{figure}

In general there is no closed analytic expression for this phase-shift
 ($\delta$), or the response amplitude of the master-sequence
 concentration.  When the fluctuations of the fitness peak is a small
 harmonic oscillation equation~\ref{eq9} becomes analytically
 tractable. For such fluctuations
\begin{eqnarray}
	\dot{y} (t) & = & 1 - (Q \sigma (t) -1) y (t)  \\
	\sigma (t) & = & \bar{\sigma} + \epsilon \sin ( \omega t )
\label{eq10}
\end{eqnarray}
From equation~\ref{eq6} it is reasonable to assume the solution to be
of the form $y (t) =\frac{1}{Q \bar{\sigma} -1} + u(t)$, where $u (t)$
is small compared to the average. Ignoring higher order terms
equation~\ref{approx} can be written in terms of the perturbation $u (t)$
as
\begin{eqnarray}
	\dot{u} (t) & = & (1 - Q \bar{\sigma}) u (t) - \frac{\epsilon Q \sin (\omega t )}{Q \bar{\sigma}-1}
\label{eq11}
\end{eqnarray}
This differential equation can be solved to obtain
\begin{eqnarray}\label{eq:12a}
	u (t) & = & - \frac{\epsilon Q}{(Q \bar{\sigma}-1) \sqrt{(Q \bar{\sigma}-1)^2 + \omega ^2}} \sin ( \omega t - \delta )  \\
	\tan ( \delta ) & = &  \frac{\omega}{Q \bar{\sigma} -1}
\label{eq12}
\end{eqnarray}
In eq.~\ref{eq:12a} and \ref{eq12} transients have been ignored since
they decay exponentially as $e^{-(Q \bar{\sigma} -1)t}$. Thus the
frequency of the oscillations is normalized by the (average) response
rate of the population $Q \bar{\sigma} -1$.

\begin{figure}[!t]
\centering
\leavevmode
\epsfxsize = 0.9 \columnwidth
\epsfbox{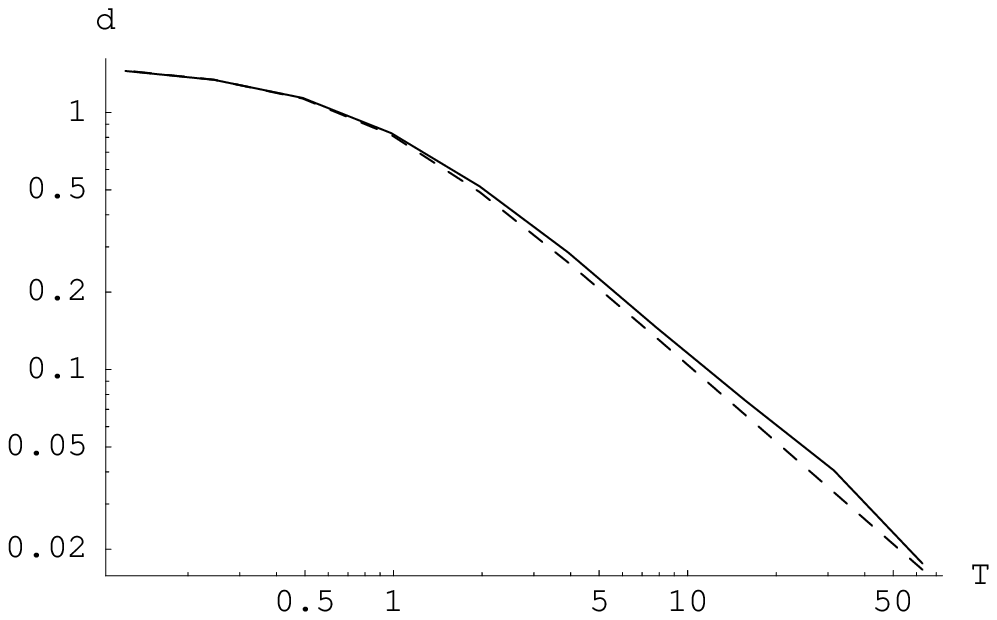}
\caption{\small  The phase shift as a function of the period $T=\frac{2 \pi}{\omega}$.
The dashed line is a prediction using Eq.~\ref{eq12} and the solid is derived by numerically
solving the reate equations~\ref{eq2}. Parameters used are $\sigma = 10 + \sin (\omega t)$, $Q = 0.7$ and $\nu = 25$.}
\label{phase}
\end{figure}

Substituting this back into the expression for $x_{0} (t)$ gives
\begin{eqnarray}
	x_{0} (t) & = &  \frac{\bar{x}}{1 -  \frac{\epsilon (1-Q) 
	\sin ( \omega t - \delta )}
	{(\bar{\sigma}-1) \sqrt{(Q \bar{\sigma}-1)^2 + \omega ^2}} }
\label{eq13}
\end{eqnarray}
where $\bar{x} = \frac{Q \bar{\sigma} - 1}{\bar{\sigma}-1}$.

The characteristic behavior of a low pass filter is clearly shown in
equation~\ref{eq12} and ~\ref{eq13}. As the frequency of the
fluctuations increases, the amplitude of the concentration response 
decreases and the phase shift converges to
$\frac{\pi}{2}$. Figure~\ref{fig:harm} shows how a population responds
to harmonic oscillations of the fitness peak. The phase-shift makes
the concentration of the mastersequence reach its maximum when the
actual fitness has already decreased below maximum.

\section{Conclusions}

In this paper we have shown that the time dynamics of a quasispecies
on a fluctuating peak can be studied under the standard no
back-mutation approximation. The general time ordering problem
stemming from a time dependent landscape disappears since the rate
equation becomes one--dimensional. We show that the time dependent
equivalent to the static error threshold is determined by the time
average of the fluctuations of the fitness peak. An expression for the
typical response time for a population is given in terms of copying
fidelity and selection pressure. We also show that for small periodic
fluctuations the time dynamics of the population has a phase shift and
a low pass filter amplitude response. Analytic expressions for the
phase shift and the amplitude are derived in the special case of small
harmonically oscillating fluctuations.

When doing this work Nigel Snoad and Martin Nilsson were supported by 
SFI core funding grants. N.S. would also like to acknowledge the 
support of Mats Nordahl at Chalmers University of Technology
while preparing this manuscript. We also Mats Nordahl for valuable comments
and discussions.

\bibliography{evolution}

\bibliographystyle{unsrt}

\end{multicols}

\end{document}